\documentclass[aps,pre,twocolumn,superscriptaddress]{revtex4}

\usepackage{graphicx,subfigure}
\usepackage{amssymb}
\usepackage{amsmath}
\usepackage{color}
\usepackage[british]{babel}

\begin{document}

\title{Deformation propagation in responsive polymer network films}

\author{Surya K. Ghosh}
\affiliation{Institute for Physics \& Astronomy, University of Potsdam,
14476 Potsdam-Golm, Germany}
\author{Andrey G. Cherstvy}
\affiliation{Institute for Physics \& Astronomy, University of Potsdam,
14476 Potsdam-Golm, Germany}
\author{Ralf Metzler}
\affiliation{Institute for Physics \& Astronomy, University of Potsdam,
14476 Potsdam-Golm, Germany}
\affiliation{Department of Physics, Tampere University of Technology, 33101
Tampere, Finland}

\date{\today}

\begin{abstract}
We study the elastic deformations in a cross-linked polymer network film triggered
by the binding of submicron particles with a sticky surface, mimicking the
interactions of viral pathogens with thin films of stimulus-responsive polymeric
materials such as hydrogels. From extensive Langevin Dynamics simulations we 
quantify how far the network deformations propagate depending on the elasticity
parameters of the network and the adhesion strength of the particles. We examine
the dynamics of the collective area shrinkage of the network and obtain some
simple relations for the associated characteristic decay lengths. A detailed
analysis elucidates how the elastic energy of the network is distributed between
stretching and compression modes in response to the particle binding. We also
examine the force-distance curves of the repulsion or attraction interactions
for a pair of sticky particles in the polymer network film as a function of the
particle-particle separation. The results of this computational study provide new
insight into collective phenomena in soft polymer network films and may, in
particular, be applied to applications for visual detection of pathogens such
as viruses via a macroscopic response of thin films of cross-linked hydrogels.
\end{abstract}

\maketitle

\section{Introduction}

Cross-linked networks of polymers of varying stiffness are a ubiquitous constituent
of biological cells, their plasma membranes, as well as the tissues they make up,
and biofilms. Such intra- and extracellular networks include, \emph{inter alia},
filaments such as microtubules, actin, collagen, fibrin, polysaccharides, or the
spectrin network maintaining the shape of red blood cells \cite{network}. In the
laboratory, cross-linked polymer gels are widely used for the separation of
molecules in setups such as gel electrophoresis \cite{gary}. Collective
deformations of such elastic networks occur, for instance, due to the motion of
individual cells in biofilms or the extracellular matrix, or when large objects
such as viruses are taken up and transported inside cells by molecular motors
\cite{endocytosis}. 

Elastic deformations and propagation of strain in networks of flexible,
semi-flexible, and stiff polymers were analysed in a number of analytical
and simulations studies \cite{frey03,head03,ever99,jam07}. Specifically, both
affine and non-affine deformations were studied \cite{affine-nonaffine}, and
the nonlinear stress-strain behaviour investigated \cite{macintosh09}. For
pre-stressed elastic networks, the rheological properties and the local mechanical
response were investigated \cite{head13}. Could one use the deformation
characteristics of polymeric networks to diagnose the presence of submicron
particles with specific surface properties, such as viruses, in the ambient
liquid?

Indeed, there exist a number of experimental techniques implementing polymeric
networks for the detection of pathogens, including setups based on highly specific
aptamers \cite{china-aptamer} or antibody-antigen interactions
\cite{virus-detection-antibody-1,virus-detection-antibody-2}. Many of these
techniques adapt techniques based on virus immobilisation on supported surfaces.
Another strategy for pathogen detection was recently proposed based on experimental
studies of polymeric hydrogels \cite{lasch-smart-surfaces}, and virus detection
based on pre-stretched DNA bundles embedded in a hydrogel was rationalised
theoretically \cite{shin13}. Hydrogel-based polymeric and polyelectrolyte materials
are known to exhibit a highly responsive behaviour with respect to various external
stimuli. The list includes the response to the ambient temperature (including
polymeric micro- and nanogels \cite{spain13nanogels,spain13nanogels-chemrev}), the
solvent quality, pH value, as well as the presence of various small molecules
\cite{responsive-hydrogels,okay09,gels-sm1,gels-sm2,gels-sm3}. The shrinkage of
polyelectrolyte microgels at elevated temperatures as triggered by
temperature-dependent (hydrophobic) interactions was recently studied by
simulations \cite{molina13shrinkageSM}. 

However, volume-based methods of pathogen detection involving hydrogel setups are
not expected to have high yields. This is due to the fact that the typical mesh
size of the gel network in many cases is comparable to or even smaller than the
size of the pathogens (even small viruses measure some 15 to 20 nm). The
diffusion of such comparatively large objects into the bulk of a hydrogel is
strongly impeded and may even exhibit transient anomalous diffusion \cite{aljash14},
and these larger particles are thus likely to predominantly occupy a small 
\emph{surface\/}
layer. One alternative is to chemically disintegrate the viral components into
nucleic acids and capsid proteins or to release easily detectable mobile compounds
from the surface of adsorbed particles that then diffuse into the bulk of a
biosensor \cite{ghosh12} which may be combined with fluorescence-based methods
\cite{neher-laschew-13}. Another alternative is to use a relatively thin, pseudo
two-dimensional supported film of an elastic polymeric material, to which pathogens
of varying sizes and adhesive strengths bind \cite{lasch-smart-surfaces}. The
physical response of this second kind of systems is the main focus of the current
study.

\begin{figure*}
\includegraphics[width=18cm]{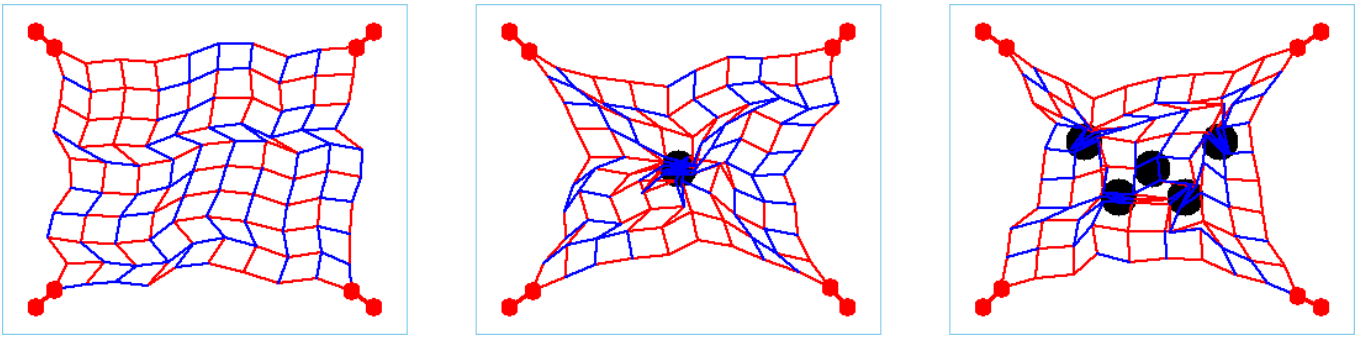}
\caption{Schematic of the responsive polymeric film supported by fixed linkers
at its four corners. Network deformations spontaneously occur due to thermal
fluctuations (left) and are enhanced in the presence of a single (middle) or
several (right) sticky particles. Stretched bonds are shown in red, while
compressed bonds appear blue. A relatively small lattice with an edge length of
$n=9$ bonds is shown to illustrate the stretching-compression features. The
Supplementary Material contains a video illustrating the dynamics of the film
shrinkage for 0, 1, and 5 sticky particles in a relatively soft network and
high attraction strength. We use the same parameters for the snapshots shown
here: attraction strength $\epsilon_A=45k_BT$ and network elasticity $k=15$, see
text. The video illustrates that the area initially shrinks rapidly, while at
later times it reaches a saturation.}
\label{fig-model}
\end{figure*}

Based on extensive Langevin Dynamics simulations we study in detail how elastic
deformations of a cross-linked polymer network are effected by introduction of
submicron particles into a sticky surface. In particular, we quantify how the
deformations propagate into the network, as function of the system parameters
given by the network elasticity and the adhesion strength of the particles. The
analysed quantities are the collective network shrinkage and the associated decay
length. We also examine the distribution of the elastic energy in the network, in
particular, its partitioning into stretching and compression modes. In addition to
effects caused by a single sticky particle we also study the network-mediated
interactions of two sticky particles.

\begin{figure*}
\includegraphics[width=18cm]{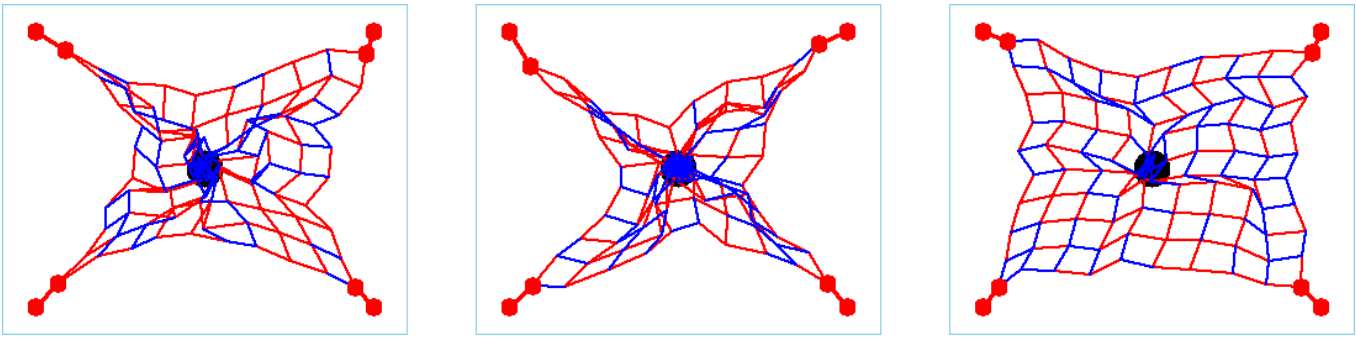}
\caption{Network response to a sticky particle analogously to
Fig.~\ref{fig-model}, for different bead-particle attraction strengths $\epsilon_A$
and network stiffness $k$. Left: $\epsilon_A=15$ and $k=15$. Middle: $\epsilon_A=
45$ and $k=15$. Right: $\epsilon_A=15$ and $k=45$.}
\label{fig-model1}
\end{figure*}

In section \ref{sec-model} we specify our simulations and define the quantities to
be analysed. In section \ref{sec-results} we present the main results of the
simulations and support them with scaling relations. Specifically, we first analyse
the binding of a single sticky particle to the thermally activated bead-spring
lattice (section \ref{sec-rad}). Section \ref{sec-partit} focuses on static
properties and determines the partition of the system, followed by results for
the elastic propagation dynamics in section \ref{sec-dyn-two}. The effects for
the network-mediated interactions between two particles in the film are examined
in section \ref{sec-int}. Finally, we draw our conclusions in section
\ref{sec-discussion}.

\section{Model}
\label{sec-model}

We use Langevin Dynamics simulations to study the elastic deformations of the two
dimensional, discretised lattice-based model of the gel film. The latter consists
of a square lattice of Lennard-Jones beads connected with elastic springs, see
Fig.~\ref{fig-model}. The elastic network with a lateral size $n$ (i.e., along
one edge we put $n$ bonds) contains $(n+1)^2$ beads and $2n(n+1)$
springs connecting them. We choose the location of the origin in the lattice
centre, and for the lattice size $n$ the $x$ and $y$ coordinates then vary from
$-\frac{n}{2}\times2a$ to $\frac{n}{2}\times2a$ with a step size $2a$. In the
simulations we vary the system size
in the range $9<n<25$, in order to eliminate boundary effects. Each bead is
subjected to the thermal bath, whose noise strength is linked to the temperature
of the system. The entire elastic film is anchored at its four corners to maintain
the shape and to prevent the total collapse of the network onto the attractive
particles, which are treated as immobile. During the simulations we keep track of
the elastic energy of the anchoring bonds, which is added to the total elastic
energy discussed below.  

One of the central quantities we target in our study is the cumulative
binding energy $E_A$ between a circular sticky particle and the beads of the
elastic network. This quantity then provides a basis to gauge the interactions
of viral particles with chemically functionalised cross-link points in the
polymeric hydrogel network. The strength of the particle-bead adhesion $\epsilon
_A$ is a model parameter, that can be tuned in experiments via, for instance, a
chemical functionalisation of the virus surface changing its affinity to a given
polymeric film. It is also sensitive to the solutions conditions, in particular,
the concentration and valency of the ambient salt.

The interaction of one of the sticky particles with the beads of the network follows
a truncated short-range 6-12 Lennard-Jones (LJ) potential of the form
\begin{equation}
\label{pot-attr}
E_{\text{attr}}(r)=4\epsilon_A\left[\left(\frac{\sigma}{r}\right)^{12}-\left(\frac{
\sigma}{r}\right)^{6}\right]+\epsilon_A,\,\,\text{for}\,\,r<2^{1/6}\sigma,
\end{equation}
and $E_{\text{attr}}(r)=0$ otherwise. The interaction radius of this attractive
potential is $r_a\approx0.2a$, recalling that the lattice constant is $2a$.
This radius controls the number of beads $N_b$ counted as bound to the sticky
particle in the simulations. Depending of the position of the bead in this
attractive potential shell surrounding the sticky particle, the energy gain due
to bead adsorption can be smaller than the maximum depth of the attractive
potential determined by $\epsilon_A>0$. Namely, the magnitude of the overall
adsorption energy $E_A$ for $N_b$ bound beads is then
\begin{equation}
\label{en_ineq}
E_A\leqslant N_b\epsilon_A.
\end{equation}
The interactions between the network beads are represented by the
standard Weeks-Chandler-Andersen repulsive LJ-like potential \cite{wca}.

The network dynamics are governed by the standard Langevin equation in the
presence of white Gaussian noise, that independently agitates each bead of the
network. If we label each network bead by the numbers $i$ and $j$ denoting its
position in the network in $x$ and $y$ direction and call $\mathbf{r}_{i,j}(t)$
this bead's position in the embedding scape, the dynamic equation reads
\begin{eqnarray}
\nonumber
m\frac{d^2\mathbf{r}_{i,j}(t)}{dt^2}&=&-\sum_{J}\nabla E_{\text{attr}}
(|\mathbf{r}_{i,j}-\mathbf{R}_{v,J}|)\\
\nonumber
&&-\sum_{m,n=1,m\neq i,n\neq j}^{(n+1)}\nabla E_{\text{LJ}}(|\mathbf{r}_{i,j}-
\mathbf{r}_{m,n}|)\\
&&-\nabla\epsilon_{el,)i,j)}-\xi\mathbf{v}_{i,j}(t)+\mathbf{F}(t).
\end{eqnarray}
Here $m$ is the mass of the bead, $\xi$ is the friction coefficient, $v_{i,j}$ is
the bead velocity, $\mathbf{R}_{v,J}$ is the position of the surface of the sticky
particle $J$, and $\mathbf{F}(t)$ represents the Gaussian $\delta$-correlated noise
with norm $\left<F(t)F(t')\right>=4k_BT\xi\delta(t-t')$.

The elastic energy stored in the springs connecting the beads at the bead-bead
separation $\delta r_{(i,j)(i+1,j)}=|\mathbf{r}_{i,j}-\mathbf{r}_{i+1,j}|$ is
parameterised by the harmonic spring potential with the equilibrium distance
$2a$,
\begin{eqnarray}
\nonumber
\epsilon_{el,(i,j)}&=&\frac{k}{2}\Big[\Big(\delta r_{(i,j)(i+1,j)}-2a\Big)^2\\
\nonumber
&&+\Big(\delta r_{(i,j)(i-1,j)}-2a\Big)^2\\
\nonumber
&&+\Big(\delta r_{(i,j)(i,j+1)}-2a\Big)^2\\
&&+\Big(\delta r_{(i,j)(i,j-1)}-2a\Big)^2\Big].
\label{pot-bond}
\end{eqnarray} 
The elastic constant $k$ varies in our simulations is the range $5\leqslant k
\leqslant135$ (in dimensionless units), and the total elastic energy of the
network is then
\begin{equation}
E_{el}=\frac{1}{2}\sum_{i,j=1}^{(n+1)}\epsilon_{el,(i,j)}.
\end{equation}
Depending on the sign of the terms $\delta r_{(i,j)(i+1,j)}-2a$, we distinguish
stretching and compression modes. 

For a typical hydrogel the measured Young's modulus $Y$ varies in a broad range
$Y=0.01\ldots20$ kPa \cite{hydrogel-E}, depending on a number of parameters such
as the mesh size and the volume density of the network.
The region of dimensionless elastic constants $k$ in our simulations which
corresponds to the Young modulus of hydrogels with the mesh size of 50 nm
accommodating typical-size virions is $635<k<5/16$. The range of $k$
in our simulations varies within this physical range for hydrogels.
Depending on the
value of $k$ with respect to the adsorption energy $\epsilon_A$, the binding of
a sticky particle yields different degrees of film deformation, compare the three
snapshots presented in Fig.~\ref{fig-model1}.
At every time step we measure the excess elastic energy, that is, the energy on
top of the basal energy level for film deformations caused solely by thermal
fluctuations, $E_{el}(t)-E_{el}^{th}(t)$. There exist several analogies to our
computational model. For instance, the response of elastic gels and the area
reduction composed of magnetic particles interconnected by a polymeric network
has recently been examined by computer simulations in Ref.~\cite{holm-sm}. The
range of elastic constants and LJ strengths used in that study is similar to our
parameter range. We implement the simplest cross-linking method of network beads
on a square lattice mimicking the often complex interconnected structures of real
hydrogels, see Fig.~7 of Ref.~\cite{new-hydrogels-sm}.

Our simulations are based on the velocity Verlet algorithm \cite{simul-book} with
integration time-step $\Delta t=0.001$. One elementary step in simulation units is
converted to $\delta\tau=a\sqrt{m/(k_BT)}$ in real time units. Using realistic
parameters for the mesh size of 50 nm in typical hydrogels
\cite{mesh-hydrogels-1,mesh-hydrogels-2},
we estimate this time as $\delta\tau\approx6$ ns. In all plots illustrating the
dynamics of the system we use the units of simulations time, $\Delta t$. The
radius of the sticky particles placed in this lattice is $R=0.9a$, which is a
fixed quantity throughout this study. Unless indicated otherwise, the figures
presented below are obtained from averaging over $M=500$ independent realisations
of the particle-network system.

\begin{figure*}
\includegraphics[width=7.6cm]{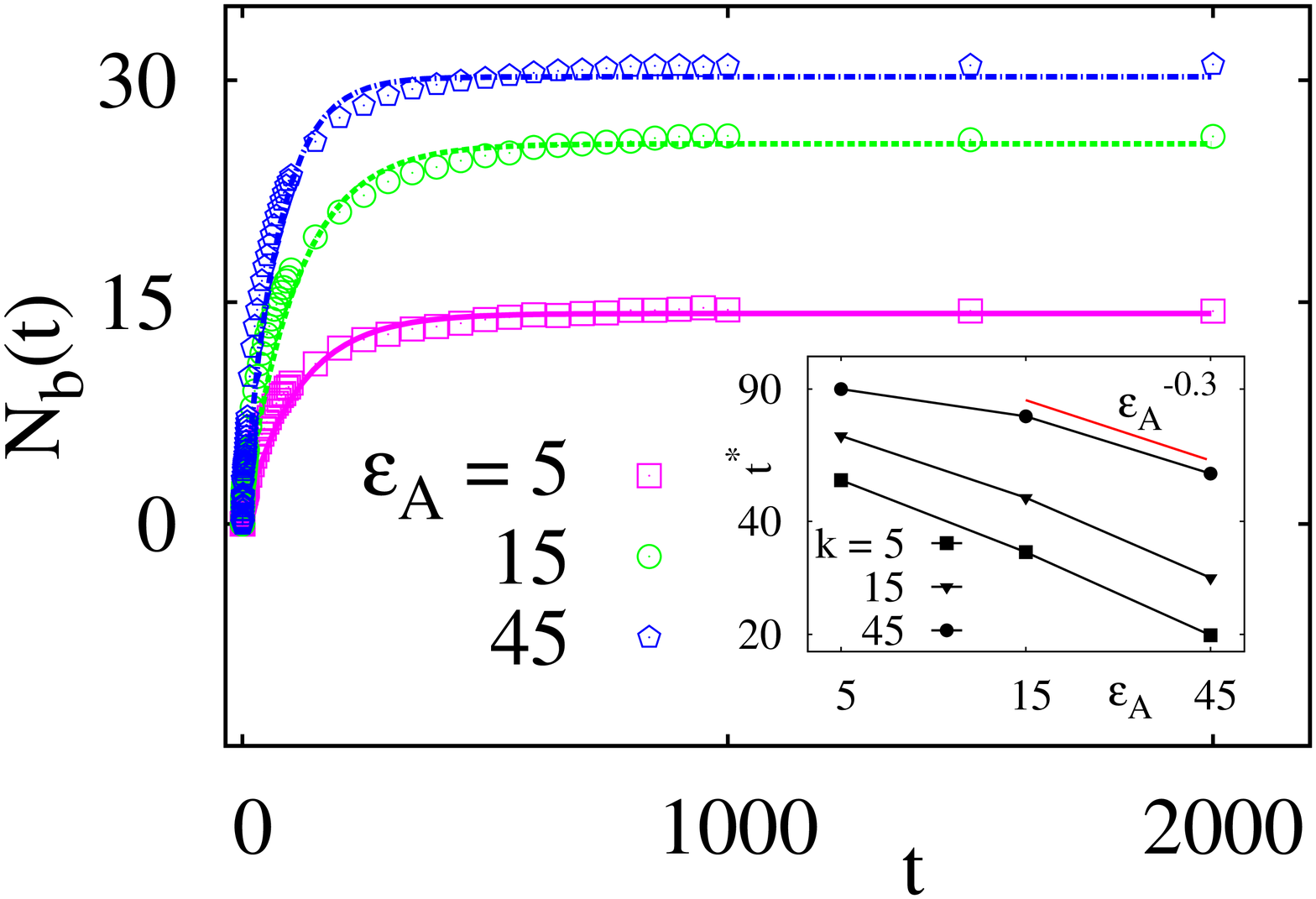}
\hspace*{0.8cm}
\includegraphics[width=7.6cm]{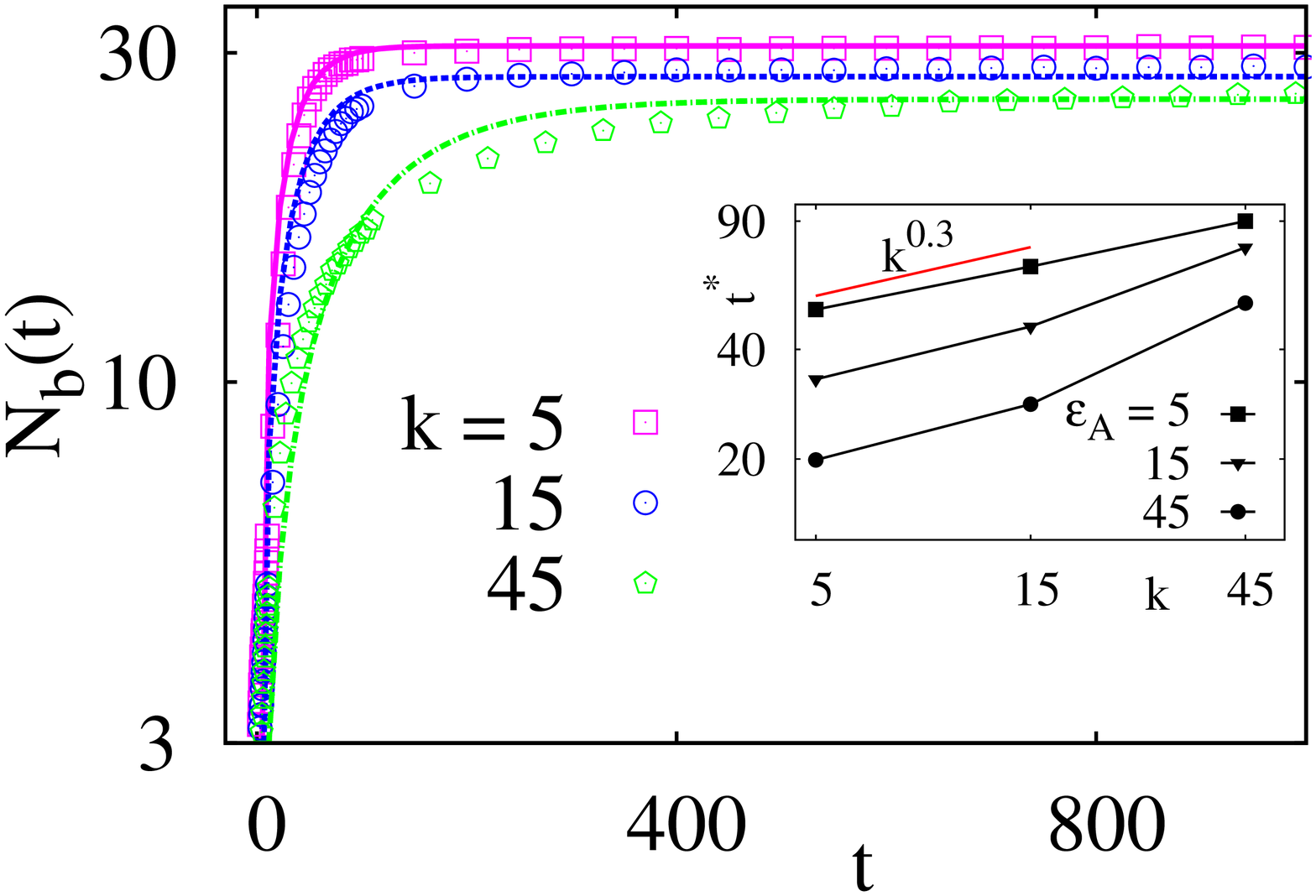}
\caption{Left: Time dependence of the number $N_b(t)$ of bound beads for different
attraction strengths $\epsilon_A$, for fixed elastic constant $k=45$ and lattice
size $n=11$. Fits to Eq.~(\ref{eq-nbound-t-fit}) are represented by the lines, and
time units are measured in number of simulations steps. $M=500$ realisations are
used for averaging.
Inset: crossover time $t^\star(\epsilon_A)$ from Eq.~(\ref{eq-nbound-t-fit}) as
function of $\epsilon_A$. Right: $N_b(t)$ on double-logarithmic scale, showing the
difference in convergence to the stationary value $N_b^{\text{max}}=\lim_{t\to
\infty}N_b(t)$, shown for varying gel stiffness $k$ and for fixed $\epsilon_A=15$.
Inset: crossover time $t^\star(k)$ as function of $k$.}
\label{fig-nb-t}
\label{fig-nb-t-k}
\end{figure*}

\section{Results}
\label{sec-results}

We present the results of our simulations, starting
with the the dynamics of a sticky particle in the network, followed by the
the static partition. We then move to the dynamic and static properties when two
sticky particles are introduced into the network.

\subsection{Dynamics: number of adsorbed beads, total film area, and elastic
energy}
\label{sec-rad}

The binding of sticky particles to the adhesive network beads alters the spectrum
of their thermal fluctuations. The strength of the particle adhesion governs the
interplay between the enthalpic and entropic contributions in the network free
energy. Namely, stronger particle binding restricts more severely the conformational
freedom of network elements. We observe that softer networks and stronger binding
energies of particles trigger more extensive stress propagation in the networks, as
expected. Note that the outer contour of the deformed polymer networks observed in
the simulations (Fig.~\ref{fig-model}) remind typical conformations of substrate
adhering biological cells \cite{cells}.    

Our simulations show that the number $N_b$ of gel lattice beads initially increases
quickly with the simulation time $t$ and then saturates to the steady state value
$N_b^{\text{max}}=\lim_{t\to\infty}N_b(t)$, as shown in Fig.~\ref{fig-nb-t}. The
number of bound
beads $N_b$ is determined according to the following criterion: if the average
bead-particle distance during the simulations is smaller than $R+r_a=R+0.2a$, it is
considered as bound. Stronger adhesion strengths $\epsilon_A$ trigger a faster
initial increase in $N_b(t)$, and naturally the plateau value $N_b^{\text{max}}$
also increases.

The initial growth of the number of bound beads can be described in terms of a
single exponential function according to
\begin{equation}
N_b(t)=N_b^{\text{max}}(1-e^{-t/t^\star}),
\label{eq-nbound-t-fit}
\end{equation}
with the characteristic crossover time $t^\star$. As demonstrated in
Fig.~\ref{fig-nb-t} on the left the quality of the fit in the linear representation
is excellent. We observe that the decay time $t^\star$ decreases for increasing
attraction strengths $\epsilon_A$ and decreasing elastic constants $k$. The
characteristic time $t^\star$ approximately follows the scaling laws
\begin{equation}
t^\star(\epsilon_A)\sim\epsilon_A^{-1/3}
\end{equation}
as function of the attraction strength $\epsilon_A$ and
\begin{equation}
t^\star(k)\sim k^{1/3}
\end{equation}
in dependence of the stiffness constant $k$, compare the insets in 
Fig.~\ref{fig-nb-t}. These scaling exponents remain almost constant with
varying network size (not shown).

\begin{figure}
\includegraphics[width=7.6cm]{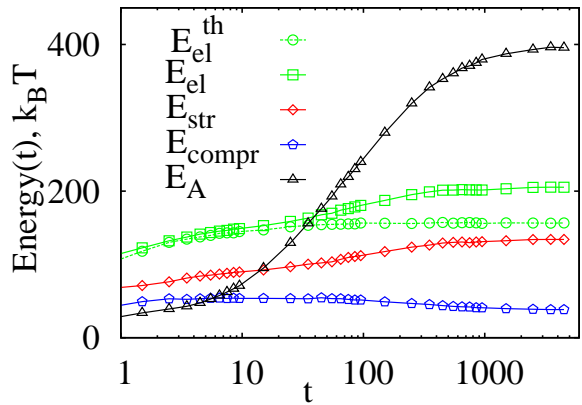}
\caption{Time dependence of the total attraction energy $E_A$ (black) between the
network beads and the sticky particle as well as the elastic energy $E_{el}$ of the
bond deformations of the gel (solid green). The latter is the sum of the stretching
$E_{str}(t)$ (red curve) and compression energies $E_{compr}(t)$ (blue). The
network elastic energy solely due to thermal fluctuations, $E_{el}^{th}(t)$, is
represented by the dashed green curve. The parameters correspond to the situation
of Fig.~\ref{fig-nb-t}: $\epsilon_A=15$, $n=11$, and $k=45$.}
\label{fig-str-compr-attr}
\end{figure}

The particle-bead attraction energy $E_A$ in the long-time limit is naturally
larger than the energy of the elastic deformations of the bonds mediated by the
particle binding, as shown in Fig.~\ref{fig-str-compr-attr}. The thermally-driven
network beads are thermally agitated and jiggle around before they get captured
by sticky particles, leading to a contraction of adjacent network springs and
subsequent compression deformations of further network elements.
The network deformation energy
is measured as the difference of the entire elastic energy minus the amount of
elastic energy caused solely by thermal fluctuations. Namely, in the long-time
limit we have
\begin{equation}
E_A>\Delta E_{el}=E_{el}-E_{el}^{th}.
\end{equation} 

The temporal increase of the adsorption and elastic energies, $E_A(t)$ and $E_{el}
(t)$, is a sensitive function of the model parameters.
The convergence to the stationary values of the attraction and elastic energies
is displayed in Fig.~\ref{fig-str-compr-attr}. We also find that larger adsorption
strengths $\epsilon_A$ and smaller elastic constants $k$ (e.g. $\epsilon_A$
= 45, $k$ = 15 as contrasted to $\epsilon_A$ = 15, $k$ = 45 considered
in the majority of plots) cause faster response of the system to particle
binding and, as a consequence, faster energy accumulation (not shown). As compared
to the thermal elastic energy that scales with the area of the system,
$E_{el}^{th}(n)\sim(n+1)^2$, the elastic deformations mediated by a weakly bound
viral particle turn out to be relatively small, $\Delta E_{el}\lesssim\
E_{el}$, see Fig.~\ref{fig-str-compr-attr}. The magnitude of $\Delta E_{el}$
grows when strongly adhesive particles are positioned in the elastic
network and the attraction energy exceeds the thermal energy $E_{el}^{th}$. Note
that relating the attraction energy presented
in Fig.~\ref{fig-str-compr-attr} to the number of bound beads shown in
Fig.~\ref{fig-nb-t} we conclude that due to binding to the
particle surface not all beads gain the maximum attraction energy given by
$\epsilon_A$, as indicated in Eq.~(\ref{en_ineq}).   

\begin{figure*}
\includegraphics[width=7.6cm]{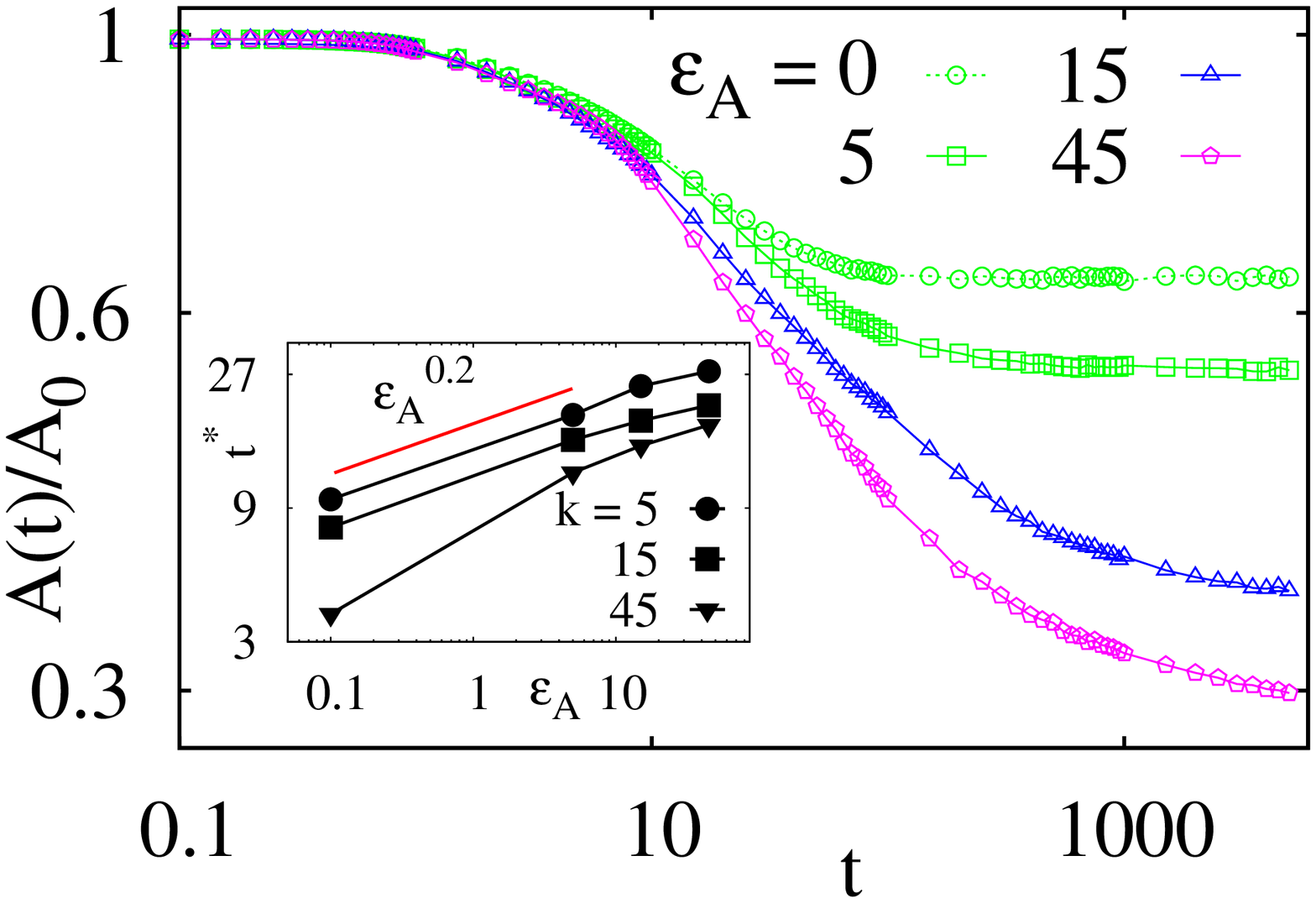}
\vspace*{0.8cm}
\includegraphics[width=7.6cm]{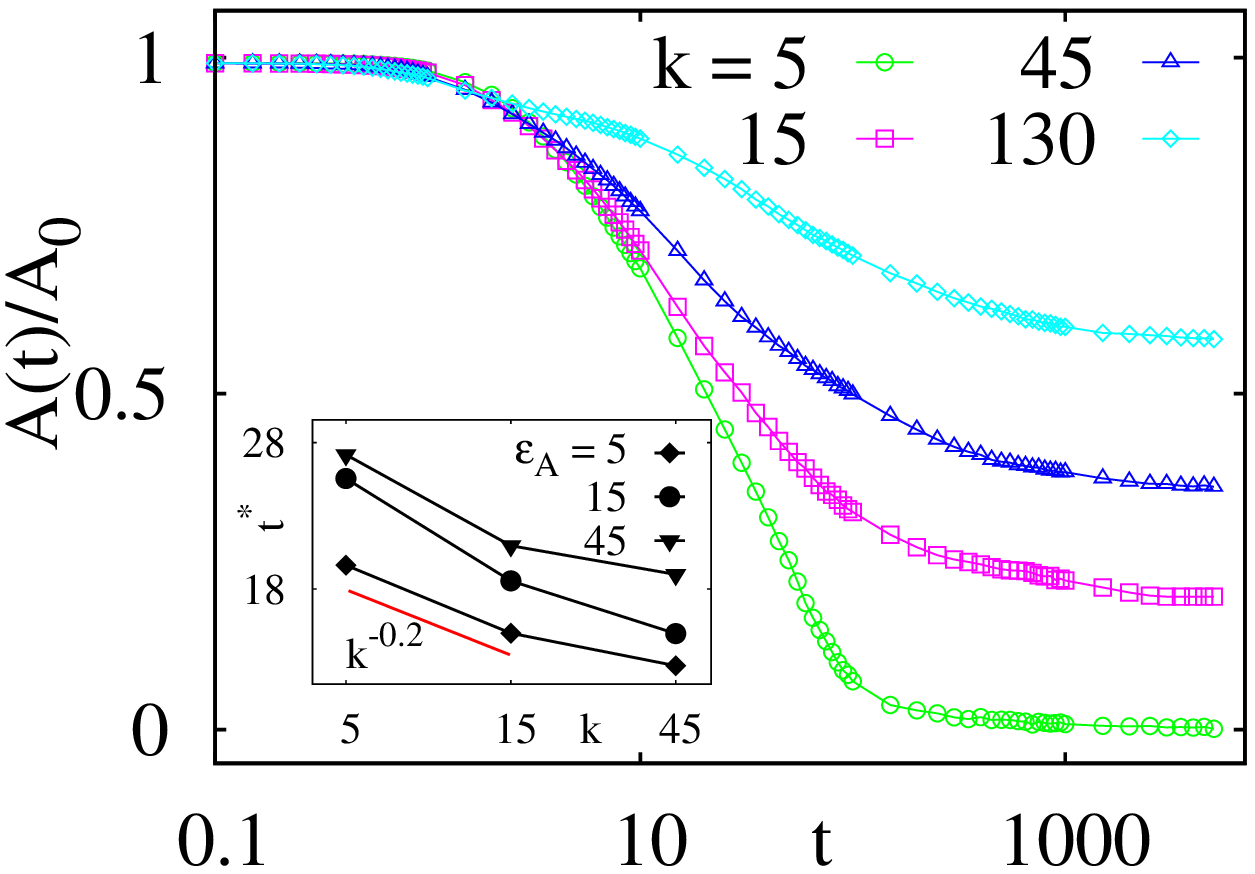}
\caption{Left: Time dependence of the relative network area $A(t)/A_0$ for different
attractions strengths $\epsilon_A$. The area shrinkage in the absence of sticky
particles is the dashed green curve. The inset illustrates the dependence of the
crossover time $t^{\ast}(\epsilon_A)$ in Eq.~(\ref{eq-area-scaling}) on
$\epsilon_A$. The points at $\epsilon_A=0$ correspond to the area relaxation of
the lattice purely due to thermal effects in absence of sticky particles. The
parameters are the same as in Fig.~\ref{fig-str-compr-attr}.
Right: Area shrinkage for varying network stiffness constant $k$. The inset shows
the crossover time $t^{\ast}(k)$.}
\label{fig-area-eps}
\label{fig-area-eps-2}
\label{fig-area-k}
\end{figure*}

Fig.~\ref{fig-area-eps} on the left illustrates the changes in the network area
$A(t)$ due to adsorption of network beads to the sticky particle. We see that with
increasing attraction strength $\epsilon_A$ the film area decreases faster and
reaches smaller stationary values as compared to a free film in response to
thermal fluctuations. We observe that the area shrinkage with time is a sensitive
function of the attraction strength $\epsilon_A$. The shrinkage dynamics can be
fitted by the relaxation function
\begin{equation}
A(t)/A_0=e^{-t/t^\ast}+C(1-e^{-t/t^\ast}),
\label{eq-area-scaling}
\end{equation}
with a single exponential, where $A_0=(2an)^2$ is the initial network area, and
the parameter $C=C(\epsilon_A,k)$ accounts for different plateau values of $A/A_0$
at long times, as function of $\epsilon_A$ and $k$. The fit to the simulations
data is particularly good when the relative area change in the course of the
system is substantial (for large $\epsilon_A$ and small $k$). To perform a
systematic fit for both the number of bound beads in Fig.~\ref{fig-nb-t} and the
area of the network in Fig.~\ref{fig-area-eps}. We considered the time span during
which 90$\%$ of the area shrinkage is completed.

The number of bound beads $N_b(t)$ reflects the
local binding characteristics and naturally grows with the strength of the
bead-particle attraction. In contrast, the area of the film is a global property,
which involves also non-affine deformations. The latter contain, for instance,
tilts and other rearrangements of rhomb-like network elements, leaving the elastic
energy of the network approximately unchanged.

The decay time $t^\ast$ of the initial relative area drop is an increasing function
of the attraction strength $\epsilon_A$, as shown in the inset of
Fig.~\ref{fig-area-eps} on the left. A stronger bead-particle attraction strength
$\epsilon_A$ causes more extensive and longer-ranged deformations of the network,
such that also more distant beads attach to the sticky particle surface at later
times. This effects a longer time scale $t^\star$. Likewise, the response of more
elastic networks is faster but the relaxation time $t^\star$ slightly grows for
smaller $k$ values, see the inset in the right panel of Fig.~\ref{fig-area-k}.
It obeys the
approximate scaling relations $t^\ast(\epsilon_A)\sim\epsilon_A^{1/5}$ and $t^\ast
(k)\sim k^{-1/5}$. We also observe that this scaling of the decay time varies only
marginally with the system size (not shown).

Finally, for varying network stiffness we observe, as expect naively, that softer
networks allow much more extensive film deformations, compare the curves in
Fig.~\ref{fig-area-k} on the right. The initial area change follows again the
exponential law, $A(t)/A_0\sim\exp(-t/t^\ast)$. The inset in Fig.~\ref{fig-area-k}
on the right illustrates the dependence of the decay time of area shrinkage on the
network elasticity.

\subsection{Steady state: energy partitioning in the gel}
\label{sec-partit}

\begin{figure}
\includegraphics[width=7.6cm]{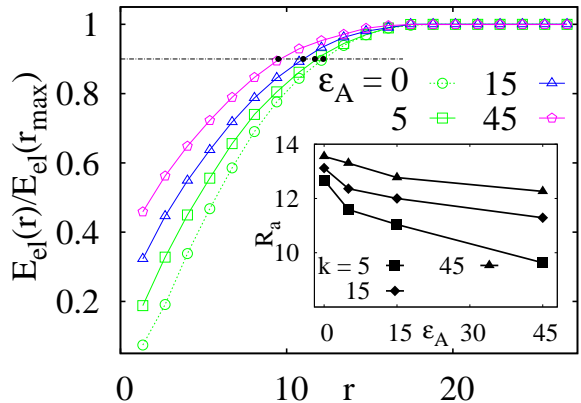}
\caption{Total elastic energy $E_{el}(r)$ as
function of the relative distance from the sticky particle in the steady state of
the lattice deformation. The radii $R_a$ of circles containing 90\% of $E_{el}$
(horizontal black dashed-dotted line in the main graph)
are shown in the inset. Energies are normalised to their maximal values achieved
at $r=r_{\mathrm{max}}$. Parameters: $n=19$, $k=15$.}
\label{fig-total-elastic-energy-accumulation}
\end{figure}

\begin{figure*}
\includegraphics[width=7.6cm]{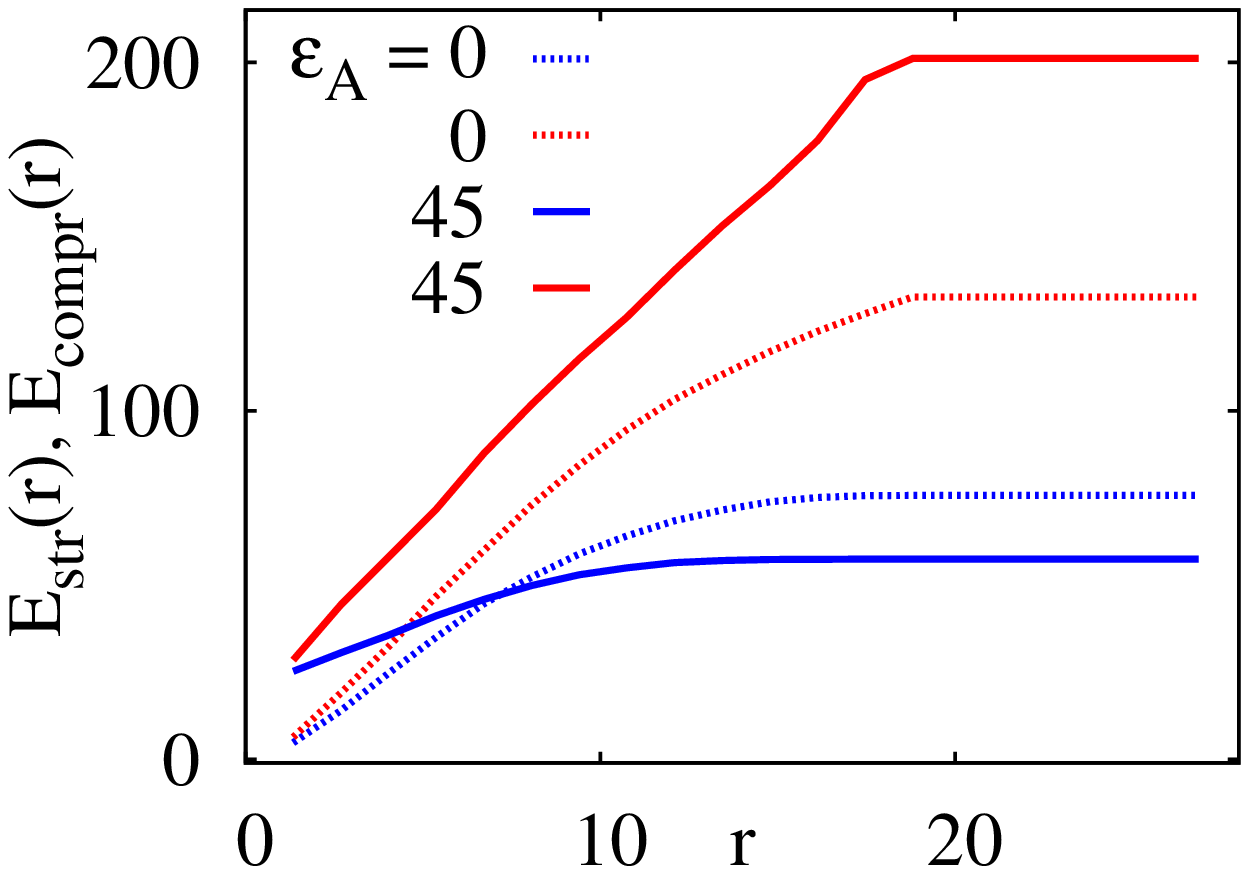}
\hspace*{0.8cm}
\includegraphics[width=7.6cm]{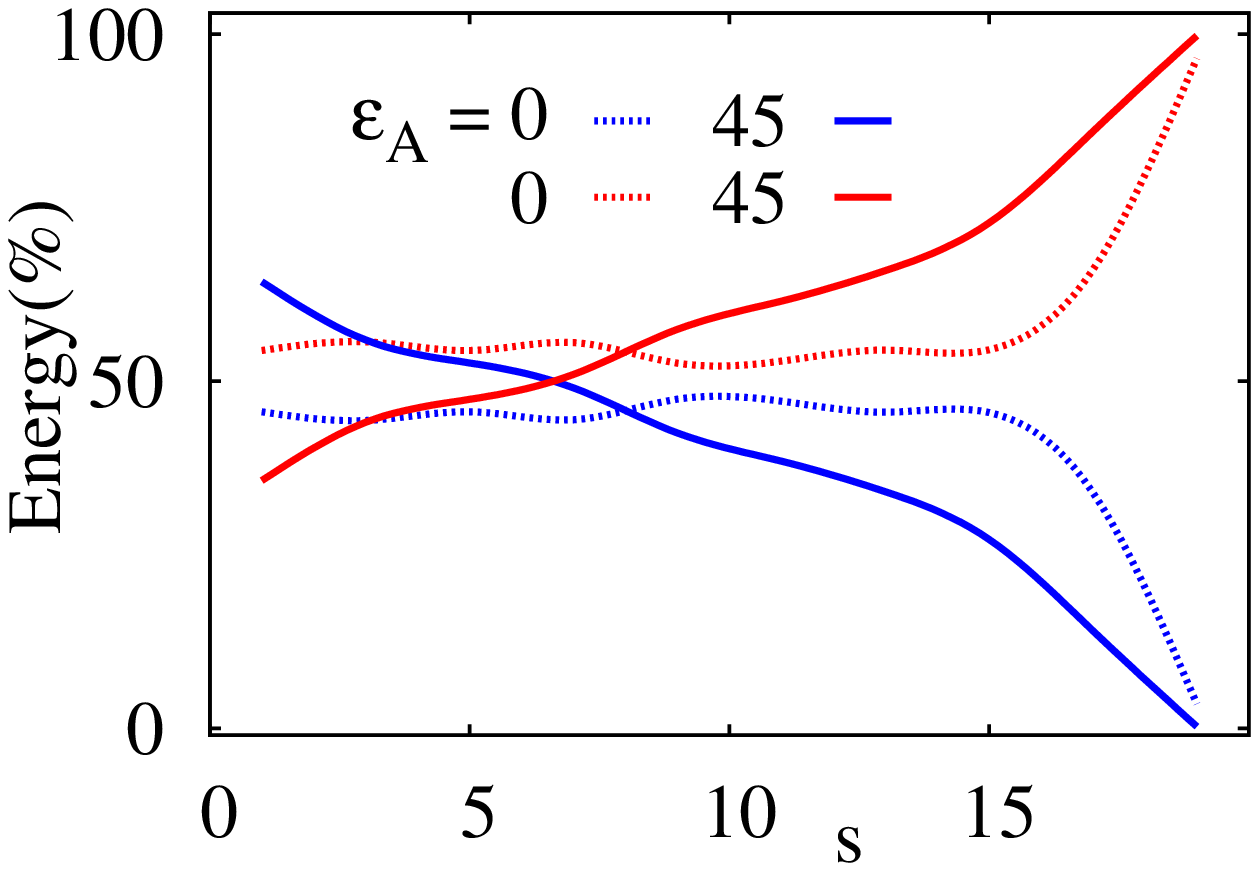}
\caption{Left: steady state distribution of stretching ($E_{str}(r)$, red) and
compression ($E_{compr}(r)$, blue) energies of the network away
from the adsorbed particle. The physical distance $r$ is the separation from the
particle in the deformed lattice, with $r_{\mathrm{max}}=2\sqrt{2}\frac{n}{2}a$.
For comparison, the  energy profiles
in the absence of bound particle $\epsilon_A=0$ are shown as well. Parameters:
$k=45,~n=19$. Right: Relative stretching (red) and compression (blue) energies as
functions of the lattice site number $1<s<n$ for the non-deformed
film. For comparison, the energy profile of the bare network is shown as the
dotted lines. The sum of the corresponding red and blue curves at each $s$ always
amounts to unity.} 
\label{fig-energy-partitioning}
\label{fig-energy-part}
\end{figure*}

The variation of the cumulative steady state elastic energy of the polymer film
is shown in Fig.~\ref{fig-total-elastic-energy-accumulation}. This total energy
is the sum of the stretching $E_{str}$ and compression $E_{compr}$ energies, which
at long times are written as
\begin{equation}
E_{el}(r)=E_{str}(r)+E_{compr}(r).
\end{equation}
As seen from Fig.~\ref{fig-total-elastic-energy-accumulation} the total
elastic energy contained within a radius $r$ away from the centre of the introduced
sticky particle grows quickly at smaller values of $r$ and then saturates. In other
words, most of the elastic energy is accumulated in the vicinity of the sticky
particle. As function of the attraction strength $\epsilon_A$ shows that higher
values of $\epsilon_A$ cause more distant network deformations in close
proximity of the adsorbed particle, see also the middle panel of
Fig.~\ref{fig-model1}. To rationalise this effect, we determine
the effective radius $R_a$ of network deformation around the centre of the sticky
particle that contains 90\% of the elastic energy. The dependence of $R_a$ on
the attraction strength $\epsilon_A$ and the network stiffness $k$ is  shown in
the inset of Fig.~\ref{fig-total-elastic-energy-accumulation}. Physically this
result implied that stronger particle-bead attractions cause more extensive
network deformations, while softer elastic networks yield more localised film
deformations (smaller $R_a$ values for smaller $k$ values), in accordance with
intuition.
      
The stretching and compression components of the elastic energy yield different
magnitudes close to the
surface of the sticky particle ("core" region) and further away ("bulk" of the
film). Specifically, in the vicinity of the core the compression energy dominates,
while on the boundary of the elastic film the stretching energy acquires larger
values. The profiles of the stretching and compression energies as function of the
radial separation $r$ from the centre of the sticky particle in the steady state
are shown in Fig.~\ref{fig-energy-partitioning} on the left. We observe that at
progressively larger separations from the adsorbed particle the energy profiles
are nearly flat meaning that the network bulk contribute only little to the
deformation energy. 

This energy partitioning is partially due to the fact that the entire system is
anchored at its four corners (Fig.~\ref{fig-model}). As the sticky particle is
smaller than the lattice spacing $2a$, the compression of the bonds takes place
in the core region to enable binding of separated network beads to the attractive
particle surface. Accompanying such a---generally non-affine---core compression,
the boundary of the system rather reveals stretching deformations, as witnessed
in Fig.~\ref{fig-energy-part}. The the anchoring points and the lattice boundary
effects also impact the interaction energy of two particles in the network, see
section \ref{sec-int}. 

Fig.~\ref{fig-energy-part} on the right illustrates the elastic energy partitioning
as computed in a layer-by-layer construction on the non-deformed lattice up to the
maximum radius $r_{\mathrm{max}}=2\sqrt{2}\frac{n}{2}a$.
We observe that for the bare, thermally agitated elastic film the
stretching and compression energies in the central region are equally distributed.
At the film boundary the entire elastic energy is due to stretching of the lattice
bonds, compare Fig.~\ref{fig-model}. In contrast, when a strongly attractive
particle is introduced, its surface compresses the network, and the disproportions
of the stretching and compression energies in the core region of the film (small
$s$ values) become apparent.

\subsection{Dynamics: network deformations around two viral particles}
\label{sec-dyn-two}

\begin{figure}
\includegraphics[width=8cm]{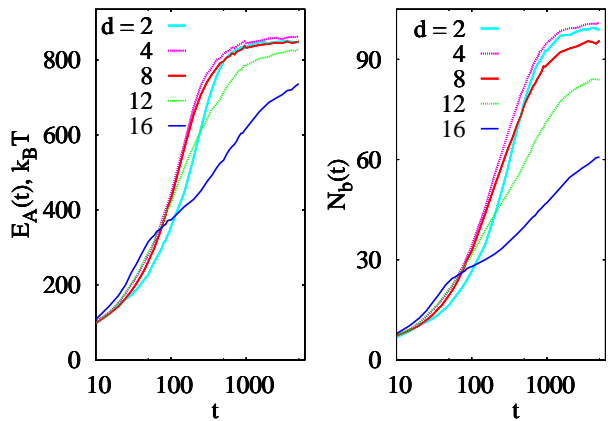}
\caption{Time dependence of the number $\mathcal{N}_b(t)$ of bound network beads
and the total adhesion energy $\mathcal{E}_A(t)$ of beads to two viral particles
introduced into the network. The particles are separated by the distance $d/(2a)$,
as indicated in the plots. Parameters: $n=25$, $k=45$, and $\epsilon_A=15$.}
\label{fig-nb-2}
\end{figure}

We now turn to study the effects of two sticky particles in the elastic network,
in particular, the particle-particle interactions. Varying the centre-to-centre
separation $d$ between these particles, we enumerate how the number $\mathcal{N}_b
(t)$ of bound network beads increases with time, as shown in Fig.~\ref{fig-nb-2}
(here and below curly letters denote two-particle quantities). We find that
for large inter-particle separations $d$ the number of bound beads is considerably
smaller than for closely-positioned particles, indicating boundary effects. 

\begin{figure}
\includegraphics[width=7cm]{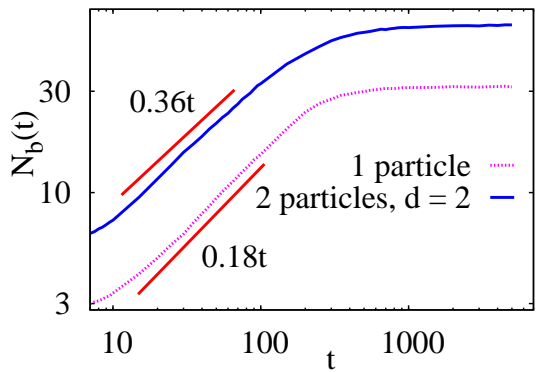}
\caption{Comparison of the number of bound beads when two sticky particles are in
the network to the case of a single particle. Parameters are the same as in
Fig.~\ref{fig-nb-2}.}
\label{fig-nb-1-vs-2}
\end{figure}

As compared to a single sticky particle introduced into the same network, the
number of bound beads increases approximately twice as fast, compare
Fig.~\ref{fig-nb-1-vs-2}. At long times, two particles are somewhat less efficient
in deforming the network compared to twice the single particle effect. Thus,
some frustration and mutual impediment of the action of two particles occurs.

A reason for the fact that the number of bound beads and the adsorption energy
are not simply proportional to one another goes as follow. According to the
adsorption criterion we defined, beads within the attraction radius are counted
as adsorbed (their favourable attraction energy is not yet in the minimum of
the potential). In the course of adsorption, the beads rearrange and their
attraction energy increases in magnitude.
At relatively large separations between the sticky
particles, the anchoring points of the film start to influence the network
response, impeding bead adsorption to the sticky particles. This is the likely
reason for the apparent kink in the time dependence of the bead adsorption energy
$\mathcal{E}_A(t)$ at short times for well separated particles as well as for
the number $\mathcal{N}_b(t)$ of bound beads, see the lowest curves at $d/(2a)=
16$ in Fig.~\ref{fig-nb-2}. We also observe that the elastic network energy for
the adsorption of two relatively weakly-adhering sticky particles, $\mathcal{E}_{
el}$, is not significantly different from that for a single adsorbed particle (not
shown). In a much larger network, obviously the contribution from two well
separated sticky particles is necessarily additive. However, for the realistic
system we have in mind, the finiteness of our simulations network reflects the
attachment of the hydrogel to its solid support in the corners.

We note that in the steady-state we observe a non-trivial dependence of the number
$\mathcal{N}_b(d)$ of adsorbed beads. Namely, for
very closely positioned particles, $d/(2a)\approx1\dots3$, the number of bound
beads decreases, indicating a shortage of the network material in between the two
particles capable of adsorption. For small separations $d$ the beads in between
the two sticky particles are frustrated in their tendency to bind to either of the
two attractive surfaces. This represents a cause for repulsive particle-particle
interactions at very close separations $d$, see below.

\subsection{Steady-state: Network-mediated particle-particle interactions}
\label{sec-int}

We finally address the question of the partitioning of the elastic energy
in the network in the presence of two sticky particles, in particular,
as function of the particle-particle separation $d$.

In this context it is worthwhile remarking that in the context of lipid membranes
the topic of membrane-mediated attractive interactions between spherical colloidal
particles \cite{attr-membr0,attr-membr,attr-membr2} and cylindrical DNA molecules
deposited on cationic lipid membranes \cite{daniel13SM-review,cp-DNA-13} attracted
significant theoretical and experimental attention in recent years. The reason for
inter-particle attraction put forward in some fluctuation-based models is due to
the tendency to reduce the inter-particle separation in order to diminish the area
of a deformed membrane in which fluctuations are suppressed due to binding of
adhesive particles. In curvature-based fluctuation-free models of membrane-mediated
attraction, in contrast, the reduction of the membrane area with large curvature
gradients in the proximity of adhered particles, that deform the membrane substrate,
causes their mutual membrane-mediated attraction \cite{cp-DNA-13}. 

To quantify the interaction energies $E_{\text{int}}(d)$ we determine the
difference of the combined adsorption ($\mathcal{E}_A(d)$) and elastic
($\mathcal{E}_{el}(d)$) energies for two sticky particles and the same
quantities for a single sticky particle in the same lattice ($E_A$ and $E_{el}$),
that is
\begin{equation}
E_{\text{int}}(d)=[\mathcal{E}_A(d)+\mathcal{E}_{el}(d)]-2(E_A+E_{el}).
\label{eq-e_int}
\end{equation}
In contrast to the membrane case, in which interaction-induced wrapping of the
membrane around adhesive particles may occur and thus the membrane assumes
out-of-plane deformations triggering the attraction, in our planar film the
network deformations are restricted to the $x$-$y$ plane. 
According to Eq.~(\ref{eq-e_int}) at large inter-particle separations $d$ the two
particles deform the film to the same extent as twice the effect of a single
particle, see also
the long time behaviour in Fig.~\ref{fig-nb-1-vs-2}. Thus, in absence of boundary
effects ($n\gg1$)
one expects the limit $E_{\text{int}}(d\to\infty)=0$. Note that finite
lattice sizes in simulations preclude this convergence. The relative contribution
of the adsorption energy versus the film deformation term in Eq.~(\ref{eq-e_int})
grows with the attraction strength $\epsilon_A$.      

\begin{figure}
\includegraphics[width=7cm]{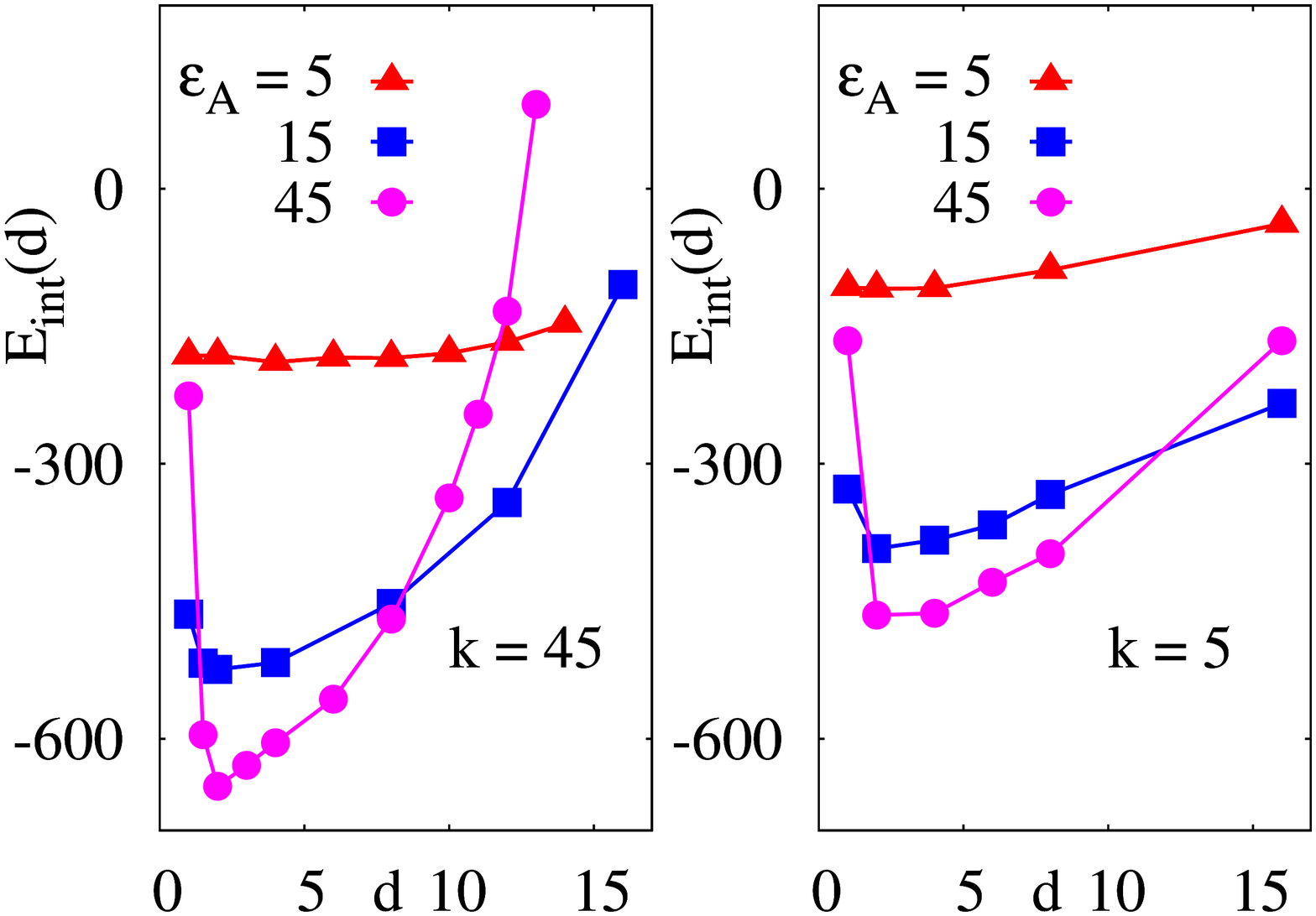}
\caption{Inter-particle network-mediated interaction energy $E_{\text{int}}(d)$
for different attraction strengths $\epsilon_A$ (left panel) and for varying
elasticity constants $k$ (right panel). The interaction energy depth grows with
$\epsilon_A$. A non-zero value of $E_{\text{int}}(d=d_{\text{max}})$ for the
largest $\epsilon_A$ indicates a growing influence of boundary effects.
Parameters: $n$=25, $k=5$.} 
\label{fig-rep-attr}
\end{figure}

We observe that the particle-particle interaction energy is non-monotonic in the
mutual separation $d$. Interactions are attractive in the region of a positive
force, that is, when $f_{\text{int}}(d)=-(\partial/\partial d)E_{\text{int}}(d)<0$,
while at close inter-particle distance $d$ the network-mediated interactions are
repulsive. At large $d$ the interaction energy increases due to a prohibited
stretching of the film boundary. In the analysis we thus should keep in mind this
finite-size effect of the film when $d/(2a)$ becomes comparable to the lattice
size $n$. Also note that at high concentrations of sticky particles introduced into
the film (right panel in Fig.~\ref{fig-model}), when the average distances are in
the region of attraction we just estimated, these particles aggregate into larger
complexes.  

The network-mediated particle-particle attraction becomes more pronounced at
higher values of the attraction strength $\epsilon_A$, compare
Fig.~\ref{fig-rep-attr}. The magnitude of the particle-particle attraction, given
by the maximum depth of the attraction energy well $E_{\text{int}}^{\text{max}}$,
naturally grows for stronger bead-particle attraction $\epsilon_A$ and decreases for
smaller network elasticity $k$, as shown in Fig.~\ref{fig-rep-attr}. A somewhat
counter-intuitive feature is the diminished attraction strength for softer networks
due to extensive binding of effectively volume-free network elements to the sticky
particles. Therefore, the
``self-energy'' of an individual particle introduced into the network is a
relatively large number, with a large portion of network beads already bound to
the particle, as one can deduce from Fig.~\ref{fig-area-k}. At the same time the
interaction energy, Eq.~(\ref{eq-e_int}), which is the relative contribution in
the particle pair as compared to the contribution of two well separated particles,
gets \textit{smaller\/} at smaller $k$ values
because of an impeded propensity of binding even more
network beads to the sticky particles, i.e., above-mentioned frustration effect.  

\section{Discussion and Outlook}
\label{sec-discussion}

We presented a simple conceptual framework to rationalise the elastic response of
a two-dimensional elastic film to the binding of sticky particles. From the
dynamical point of view, the kinetics of area shrinkage due to progressive binding
of network elements was quantified. In particular, we obtained simple scaling
relations for the characteristic relaxation times for the number of bound network
beads and the film area, as functions of the network stiffness and the attraction
strength of the particles to the network beads. In the steady-state, we obtained 
the distribution of the elastic energy as function of the distance from the
adsorbed particle and also quantified the radius of propagation of the elastic
deformations. Finally, we enumerated the interaction energies and forces acting
between sticky particles in the network which emerge due to the elastic network
deformations.

The observed strengthening of the effective particle-particle attraction for
stronger bead-particle attraction strength is similar to trends
for membrane-mediated attractions between parallel DNA molecules adsorbed on
unsupported cationic lipid membranes which wrap around the negatively charged
DNAs \cite{cp-DNA-13}. An every-day life analogy for this network-mediated
attraction between the sticky particles as well as for the distribution of
network deformations in the vicinity of a pathogen can in fact be the example of
two persons on a trampoline. In fact, this simple analogy is particularly close
when the trampoline's fabric is only anchored in its four corners.

We also note that the interactions between sticky particles deposited on soft
responsive networks are akin to capillary forces acting between particles
immersed on fluid interfaces \cite{capil-sm1,capil-sm2,capil-sm3}. The
inter-particle attraction and clustering emerges as a consequence of the
reduction of regions of network deformations with high elastic energy gradients,
focused in the proximity of adsorbed particles.

As mentioned in the introduction, the results we obtained here may open new
perspectives for the detection of particles such as viral pathogens of different
binding specifics by help of thin highly-responsive polymeric films.
The current area-shrinkage setup for viral particle detection thus
complements the model of detection we proposed recently, which
converts the virus-DNA binding events and accompanying DNA melting
into measurable volumetric changes of the hydrogel specimen \cite{shin13}.
To exploit
the whole phase space for such purposes, future simulations should include
particles with asymmetric geometries (ellipsoidal or rod-like shapes, for
instance \cite{attr-membr3}) introduced into a three-dimensional inter-connected
network, instead of the simple 2D planar case examined here. In this situation,
anisotropic deformations of the network and anisotropic inter-particle
interactions will occur. These are the targets for future investigations.

\acknowledgments

The authors thank J.~Shin and A.~Laschewsky for discussions. We acknowledge
funding from the Academy of Finland (FiDiPro scheme to RM), the German Research
Council (DFG Grant CH 707/5-1 to AGC), and the Federal Ministry of Education and
Research (SKG).

\end{document}